\documentstyle[12pt,epsf]{article}
\newcommand{\commnt}[1]{}
\newcommand{\cmmnt}[1]{}
\newcommand{\prg}[1]{}

\newcommand{\degree} {{\rm deg}}

\newcommand{\lbni}{\(<\)}
\newcommand{\rbni}{\(>\)}

\newtheorem{property}{Property}[section]
\newcommand{\symg}[1]{S_{|#1|}}
\newcommand{\Proof}[1]{{\small\bf Proof.} #1 $\Box$.}

\begin{document}
\thispagestyle{empty}
\rightline{KEK-CP-020}
\rightline{KEK Preprint 94-83}
\rightline{MGU-CS/94-01}
\rightline{August, 1994}
\vspace{1em}
\begin{center}
{\Large\bf
A Feynman-graph generator for any order of coupling constants}\\
\vspace{3em}
{\large {Toshiaki KANEKO%
\footnote{e-mail address: {\tt kaneko@minami.kek.jp}}}}%
\footnote{Present address: 
LAPP, B.P.110, F-74941 Annecy-le-Vieux CEDEX, France}\\
\vspace{1em}
{\bf\sl Meiji-Gakuin University\\
        Kamikurata-cho 1518, Totsuka, Yokohama 244, Japan\\
        {\rm and}\\
        National Laboratory for High Energy Physics (KEK)\\
        Tsukuba, Ibaraki 305, Japan
}
\end{center}

\vspace{2em}

\begin{quote}{\small
\begin{center}{\bf abstract}\end{center}
    A computer program has been developed which generates 
Feynman graphs automatically for scattering and decay processes 
in non-Abelian gauge theory of high-energy physics.
    A new acceleration method is presented for both generating and
eliminating graphs.
    This method has been shown to work quite efficiently for any order
of coupling constants in any kind of theoretical model.
    A utility program is also available for drawing generated graphs.
These programs consist of the most basic parts of the GRACE system,
which is now used to automatically calculate tree and one-loop processes.
}\end{quote}

\vspace{1em}
\begin{center}{\small
Published in Computer Physics Communications {\bf 92} (1995) 127 -- 152
}\end{center}

\vspace{2em}
\section{Introduction}

\prg{Automatic system}
    In regard to recent projects concerning experimental high-energy
physics with increasing available energy,
accurate theoretical analysis has been required for increasing the number
of scattering processes.
    Since electro-weak theory combined with QCD, known as
the standard model, is considered to be
a basic theory among various theoretical models,
exact perturbative calculations within the framework of this model are
considered to be the standard for theoretical predictions.
    Because of the complexity of interactions, the number of 
Feynman graphs for one process becomes much greater than that in QED.
    There appear several tens to hundreds of graphs in the tree
process and several hundreds to thousands in the one-loop process.
    In addition, in contrast to pure QCD, one cannot ignore various mass
parameters which play important roles.
    These situations make a theoretical calculation of the amplitude
more difficult and tedious.
    At present, it seems that the amount of labor necessary for exact
calculations has almost reached the limit of hand calculations.
 It is thus natural to utilize computers to carry out such work.
 Several groups have started independently to develop computer
systems which automate the perturbative calculations in
the standard model\cite{grace,gracel,comphep,feyncalc,madgraph}.
    Automatic calculations of the tree processes and some part
of the one-loop processes have already been achieved by these systems.

\prg{Generation of Feynman Graphs}
    An automatic calculating system starts with the
generation of Feynman graphs for a given physical process.
    Feynman-graph generation for all orders in QED is not difficult,
since the problem can be reduced to a problem equivalent
to tree-graph generation\cite{sasaki,grand}.
    For electro-weak theory, or even in the \(\phi^3\) model, however,
the situation is different.
    They include self-interactions of particles, whose vertices
are symmetric under exchanging interacting particles.
    This symmetry of the vertices results in a complicated structure of 
the internal symmetry in a Feynman graph.

    It is not difficult to generate a sufficient set of Feynman graphs
by simply connecting the vertices.
    The problem, however, is to eliminate duplicated graphs.
    It is hard to analyze the structure of the internal symmetry
of graphs so as to avoid duplicated graphs
controlling the graph generation process.
    Usually, a newly generated graph is discarded when it is found to
be topologically equivalent to one already generated.
    Although explicit comparisons of graphs are, of course, possible,
any known algorithm consumes time, which increases as an
exponential of the number of vertices.
    A graph comparison is generally not considered in the
category of {\sl polynomial time complexity}\cite{compare}.
    There is another problem in a explicit comparison method.
    It requires to save all generated graphs in order to recall
graphs for comparisons.
    In a practical problem, the list of generated graphs grows too long 
to keep in the main memory of a computer.
    So the list is recorded on a secondary memory, usually on disk space.
    Frequent access to a secondary memory makes program
very slow.
    This problem is solved by an {\sl orderly algorithm}\cite{orderly}.
    An orderly algorithm judges whether a newly generated 
graph is necessary to be kept or not, looking only at the graph
without recalling already generated graphs.
    Based on this algorithm, a Feynman-graph generator for
a wide variety of models was developed by P.Nogueira\cite{nogueira}.

\prg{Aim}
    The aim of this paper is to present a computer program for 
Feynman-graph generation that is sufficiently fast for practical use.
    For this purpose, we have developed a new method of graph
generation which accelerates the orderly algorithm.
    We classify vertices in such a way that there is no topologically
equivalent vertices in different classes.
    With this classification, we are able to reduce the number of
graphs to which the orderly algorithm is applied.
    The classification method of vertices which we use is an
empirical one developed by graph theorists in order to determine
the equivalence of two given graphs\cite{kucera}.
    We have modified this method so as to be suitable for
graph generation.

\prg{Condition of graph generation}
In order to make the problem simpler, we assume the following
conditions:
\begin{enumerate}
\item Vacuum-to-vacuum graphs are not considered.
\item External particles of the graphs are assumed to be topologically
      different.
\end{enumerate}
    The first condition stems from the fact that the purpose of our
automatic system is to calculate physical processes, namely the 
scattering and decay of particles.
    With the second condition, we have a one-to-one correspondence
between the Feynman graphs and the Feynman amplitudes.
    One can generalize the method to loosen these restrictions
without any essential difficulty, since the orderly and classification
algorithms are still applicable.
    However, it must make the program more complicated.
This generalization is discussed in section \ref{summary}.

\prg{Structure of this paper}
    In the next section we introduce our basic methods.
    An explanation concerning the orderly algorithm is also given.

    In order to accelerate orderly algorithm, we classify vertices
in section \ref{sec:class}.
    A pre-selection rule of generated graph, which is necessary
to make the orderly algorithm consistent with vertex classification,
is presented in this section.

    The method of graph generation is described in section \ref{sec:gen}.
    The vertex-classification algorithm has been combined in order to
eliminate graph in intermediate steps.

    In section \ref{impl} we introduce our implementation of
algorithms and utility programs.

    A summary and comments are given about the physical processes 
calculated by the GRACE system up to now.
    A generalization of our program without the two conditions 
mentioned above is also discussed.

    Appendices \ref{manual}, \ref{process}, \ref{model} and
\ref{grf} comprise a brief manual of the programs
and file formats.
\section{Basic method}
\label{sec:basic}

    A {\sl graph} consists of a finite number of {\sl nodes}
and {\sl edges}.
    A {\sl node} is either an external particle or a vertex.
    An {\sl edge} is either a propagator
or a connection between an external particle and a vertex.
    Throughout this paper, \(N\) denotes a set of nodes
and \(E \subset \{ (u, v) | u, v \in N \}\) does a set of edges.
    A graph \(G = (N, E)\) is defined mathematically as a pair
of these sets.
    The {\sl degree} \(\degree(v)\) of node \(v\) is defined as the number
of edges attached to the node \(v\):
\begin{eqnarray}
  \degree(v) = | \{ (u, v) \in E \} |
\end{eqnarray}
    The degree of an external particle is 1.
    An edge has several attributes in a Feynman graph, such as
the number of multiple connections between nodes and the 
name of a particle of a propagator.
    In order to make our discussion simpler, we drop these attributes
in the following description.
    A node is labeled by a positive integer, with which we identify
a node.
    We regard a set of integers \(\{ 1, ..., |N| \}\) as the
set of nodes \(N\).

    Let us consider a graph \(G = (N, E)\) and a permutation \(p\)
mapping the set of nodes \(N\) onto itself.
    The permutation \(p\) is generalized to act on the set
of edges by defining \( p (u, v) = (p u, p v)\) for \((u, v) \in E\),
\(u, v \in N\), and is generalized to act on the graph \(G\) as
\(p G = p (N, E) = (p N, p E) = (N, p E)\).
    Two graphs, \(G = (N, E)\) and \(G' = (N, E')\), are
{\sl isomorphic} when there exists such a permutation \(p\)
that maps the graph \(G\) to \(G' = p G\).
    This condition is equivalent to
\begin{eqnarray}
    (u, v) \in E \Longleftrightarrow (p u, p v) \in E'.
\label{isomorphic}
\end{eqnarray}
Isomorphic graphs are different each other only in the way
of labeling nodes.

    A permutation \(p\) is an {\sl automorphism} of a graph
\(G\) when \(p G = G\).
    The set of all automorphisms of a graph \(G\) forms a group
\(\Gamma_G\) called an {\sl automorphism group}.

\begin{figure*}[htb]
\centerline{\epsfbox{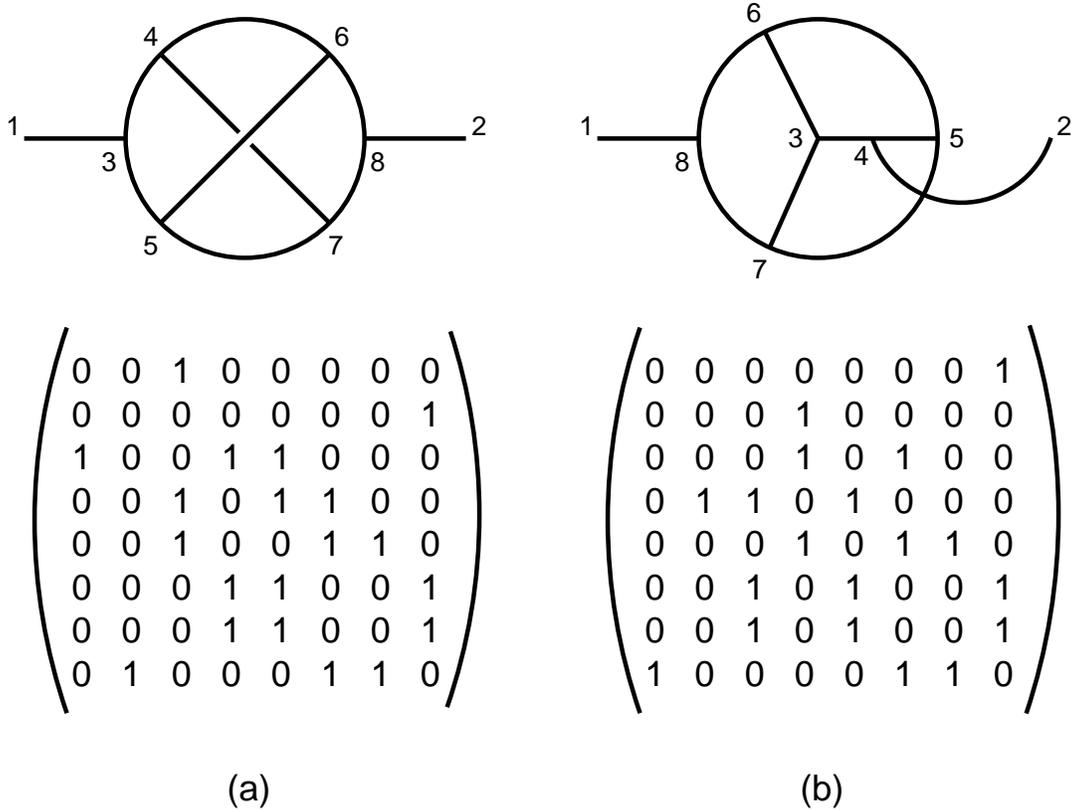}}
\caption{Example of adjacency matrix.  \label{fig:isomor}}
\end{figure*}

A graph can be expressed as a matrix, called {\sl adjacency matrix},
whose element \(a_{ij}\) carries the number of connections
between nodes \(i\) and \(j\).
We show an example of isomorphic graphs and their adjacency
matrix in Fig. \ref{fig:isomor}.

    In this and the next sections we consider a simple method of graph 
generation.
    First, necessary number of nodes are generated and then
the nodes are connected in all possible ways.
    This method creates all possible graphs.
    Topologically equivalent graphs may appear several times.
    In order to eliminate duplicate graphs, 
we use a systematic method, 
an {\sl orderly algorithm}\cite{orderly}, which judges a newly
generated graph being necessary or not without recalling any of
already generated graphs.

    An {\sl orderly algorithm} compares graphs through values of
a function, called {\sl coding}, which maps a graph to an integer 
in a way that different isomorphic graphs 
are mapped to different values.
    Such a function is easily realized,
for example, in regarding the elements of the
adjacency matrix as a sequence of digits of an integer\cite{orderly}.

    In order to pick up a representative from a set of isomorphic
graphs,
An orderly algorithm selects a graph \(G\) when it satisfies 
the following condition with coding \(f\):
\begin{eqnarray}
    f(G) = \max_{p \in \symg{N}} f(p G),
\label{maxcode}
\end{eqnarray}
where \(\symg{N}\) is the symmetry group of the set of nodes
\(N\) (the set of all permutations acting on \(N\)).
    A graph \(G\) is indirectly compared with graph \(p G\) through
their values of the coding \(f\).

    Combining this selection condition with graph generation
process, an orderly algorithm generates graphs in the following 
procedure:
\begin{itemize}
\item[1] Generate all the possible graphs.
\item[2] Apply all possible permutations to each graph.
\item[3] Discard a graph when it has a smaller coding value 
         than ones of permuted graphs.
\end{itemize}

    This method requires to keep only two graphs, a tested graph
and its permuted one, which are small enough to keep in the
main memory of a computer.

    When a permutation \(p\) satisfies selection condition (\ref{maxcode}),
\(p\) is an element of automorphism group \(\Gamma_G\) of G.
    One can construct \(\Gamma_G\) explicitly in collecting
all of such permutations.
    The symmetric factor \(| \Gamma_G |\) necessary for calculating
Feynman amplitudes is evaluated in this way.

    A simple application of this algorithm requires comparisons
among graphs corresponding to
\(O(|N|!)\) permutations for each of \(O(|N|!)\) isomorphic graphs.
    In order to decrease the number of comparisons 
in the orderly algorithm, we consider to replace \(\symg{N}\) in
(\ref{maxcode}) by its subgroup.

\section{Vertex classification}
\label{sec:class}

    We consider a classification \(\{N_i\}\) (\(\cup_i N_i = N\),
\(N_i \cap N_j = \emptyset\) for \(i \neq j\))
of the set of nodes \(N\) such that any two nodes in different
classes are topologically different.
    We do not require that nodes in a class are 
topologically equivalent.
    In other words, each class of nodes is a direct sum of
the orbits in \(\Gamma_G\) (an {\sl orbit} of a node \(v\)
in \(\Gamma_G\) is the set
\(\{ u | \exists p \in \Gamma_G, p v = u \}\)).
    We call such a classification a {\sl consistent classification}.
    With this classification of nodes, we construct a group
\begin{eqnarray}
S(\{N_i\}) &=& \symg{N_1} \otimes \symg{N_2} \otimes ...
\end{eqnarray}
where \(\symg{N_i}\) is the symmetry group acting on the set 
\(N_i\).
    We replace the group \(\symg{N}\) in condition (\ref{maxcode})
by this subgroup \(S(\{N_i\})\).
    This replacement reduces
the number of permutations from
\(|N|!\) to \(|N_1|! \times |N_2|! \times ... \).
    Corresponding to this limitation of comparisons in the 
orderly algorithm,
one must limit the set of generated graphs.
    We present the way of limiting graphs after
describing our method of node classification.

    We adopt a method of node classification used 
in graph theory\cite{kucera}.
    First, nodes are classified into \(\{N_i^{(0)}\}\) by simple
topological properties.
    Since we consider that external particles are topologically different
from each other, they are put into different classes.
    The i-th external particle is put to the i-th class.
    Other nodes are classified by their degree such that:
\begin{eqnarray}
    u, v \in N_i^{(0)} \Longleftrightarrow \degree(u) = \degree(v) \nonumber\\
    u \in N_i^{(0)}, v \in N_j^{(0)}, deg(u) < deg(v)
        \Rightarrow i < j
\end{eqnarray}
    We call this initial classification \(\{N_i^{(0)}\}\) a
{\sl primitive classification}.
    Primitive classifications \(\{N_i^{(0)}\}\)
and \(\{M_i^{(0)}\}\) of isomorphic graphs \(G\) and \(p G\), respectively,
satisfy:
\begin{eqnarray}
p N_i^{(0)} = M_i^{(0)} \mbox{\qquad for all }i, \label{iso:init}
\end{eqnarray}
since relabeling of nodes does not change topological
properties of nodes such as \(\degree(pv) = \degree(v)\).

    Starting from the primitive classification, the classes 
\(\{N_i^{(k)}\}\) are refined to \(\{N_i^{(k+1)}\}\) iteratively.
    We consider a vector \(a_v[i], i = 0, 1, ... \) for a node
\(v \in N_j^{(k)}\).
    The zeroth element \(a_v[0]\) of the vector keeps its current
class number, and the \(i\)-th element keeps the number 
of edges which connect \(v\) with nodes 
in the \(i\)-th class \(N_i^{(k)}\):
\begin{eqnarray}
a_v[0] &=& j, \mbox{\qquad for } v \in N_j^{(k)}, \nonumber\\
a_v[i] &=& |\{(u, v) \in E | u \in N_i^{(k)}\}| \mbox{\qquad for }  i > 0.
		 \label{avect}
\end{eqnarray}
The nodes are classified by the values of these vectors.
New classes are numbered in increasing order, which is evaluated in
the lexicographical ordering ``\(\prec\)'' 
of vectors. 
This process is repeated until classes can no more be refined.
Thus the nodes are classified only by their 
topological properties.
We summarize this refining method of classification as follows:
\renewcommand{\arraystretch}{1}
\begin{quote}
{\tt \begin{tabbing}
xxxx\=xxxx\=xxxx\=xxxx\=xxxx\=\kill
refine(\(\{N_i^{(k)}\}\))\\
\{\\
\>   for all ({\rm class }\(j\)) \{\\
\>   \>   for all (\(v \in N_j^{(k)}\)) \{\\
\>   \>   \>   \(a_v[0]\) = \(j\);\\
\>   \>   \>   for all ({\rm class }\(i\))\\
\>   \>   \>   \>    \(a_v[i]\) =
	    \(\bigl| \{(v, w) \in E | w \in N_i^{(k)}\} \bigr|\);\\
\>   \>   \}\\
\>   \}\\
\>   ({\rm construct \(\{N_i^{(k+1)}\}\) such that}\\
\>   \>   \(a_v     = a_w, v \in N_i^{(k+1)}, w \in N_j^{(k+1)}
		       \Rightarrow i = j\);\\
\>   \>   \(a_v \prec a_w, v \in N_i^{(k+1)}, w \in N_j^{(k+1)}
		       \Rightarrow i < j\);\\
\>   );\\
\>   return \(\{N_i^{(k+1)}\}\);\\
\}
\end{tabbing}}
\end{quote}
\renewcommand{\arraystretch}{2}
We call a sequence of this classification of nodes 
\(\{N_i^{(k)}\}\) ( \(k = 0, ..., K\) ) a {\sl refinement sequence}.

We show the following properties for a refinement sequence.

\begin{property} \label{pr:incl}
For any \(N_i^{(k+1)}\)
there exists an input class \(N_j^{(k)}\) includes \(N_i^{(k+1)}\):
\begin{eqnarray}
N_i^{(k+1)} \subset N_j^{(k)}.
\end{eqnarray}
\end{property}
This property states that \(N_i^{(k+1)}\) is a refined class
of \(N_j^{(k)}\).

\begin{property} \label{pr:ord}
Ordering of classes are kept by the refinement procedure:
\begin{eqnarray}
\forall i \; \forall j\
[
N_i^{(k+1)} \subset N_l^{(k)},
N_j^{(k+1)} \subset N_m^{(k)}, i < j
    \Rightarrow l \leq m
].
\label{subord}
\end{eqnarray}
\end{property}
This property is evident from the fact that the algorithm
renumbers the new classes according to the lexicographical ordering of 
the vector \(a_v\), which keeps the old class number at the first 
element.

Thus classes \(N_i^{(k)}\) are divided to:
\begin{eqnarray}
N_1^{(k)} &=& N_1^{(k+1)} \cup N_2^{(k+1)} \cup ... 
      \cup N_{i_1}^{(k+1)} \nonumber\\
N_2^{(k)} &=&
N_{i_1+1}^{(k+1)} \cup
N_{i_1+2}^{(k+1)} \cup
... \cup
N_{i_1+i_2}^{(k+1)} \label{kclass}\\
  & & ...\nonumber 
\end{eqnarray}

\begin{property} \label{pr:index}
When nodes of two isomorphic graphs \(G = (N, E^G)\) and 
\(H = p G = (N, E^H)\), \(p \in \symg{N}\), 
are classified into 
\(\{N_i^{(k)}\}\) and \(\{M_i^{(k)}\}\),
respectively, the following relation holds for all \(k\):

\begin{eqnarray}
\forall i \; [ p N_i^{(k)} = M_i^{(k)} ]
\Rightarrow
\forall j \; [ p N_j^{(k+1)} = M_j^{(k+1)} ]
\end{eqnarray}
\end{property}
\Proof{
Let us take a node \(v \in N_i^{(k)}\).
We consider vectors \(a_v^G\) and \(a_{pv}^H\) in eq. (\ref{avect}) 
which are used to construct \(\{N_j^{(k+1)}\}\) and 
\(\{M_j^{(k+1)}\}\) from \(\{N_j^{(k)}\}\) and \(\{M_j^{(k)}\}\), 
respectively.
Their zero-th elements are equal \(a_v^G[0] = a_{p v}^H[0] = i\)
since \(pv \in M_i^{(k)}\).
We obtain for \(l > 0\):
\begin{eqnarray}
  a_v^G[l] & = & | \{ (v, w) \in E^{G} | w \in N_l^{(k)} \} | 
                 \nonumber\\
           & = & | \{ (p v, p w) \in p E^{G} | p w \in p N_l^{(k)} \} | 
                 \nonumber\\
           & = & | \{ (p v, p w) \in E^{H} | p w \in M_l^{(k)} \} | \\
           & = & a_{p v}^{H}[l]
                 \nonumber
\end{eqnarray}
    Thus equation \(a_v^{G} = a_{p v}^{H}\) is hold 
for any node \(v\).
    Since the numbering of refined classes are determined only 
by these vectors, it is easy to see
\begin{eqnarray}
v \in N_j^{(k+1)} \Longleftrightarrow p v \in M_J^{(k+1)},
\end{eqnarray}
which implies \(p N_j^{(k+1)} = M_j^{(k+1)}\).
} 

    Since \(p N_i^{(0)} = M_i^{(0)}\) 
is satisfied for all \(i\) in eq. (\ref{iso:init}), 
the above property result in:
\begin{eqnarray}
p N_i^{(k)} = M_i^{(k)} \mbox{\qquad for all }i \mbox{ and }k.
                          \label{iso:seq}
\end{eqnarray}
    Especially, the length of refining sequences of two 
isomorphic graphs are equal.

    Now we consider the way of limiting graphs to be 
applied to the orderly algorithm.
    We impose the following pre-selection condition for 
the classification \(\{N_i\}\) of nodes of a generated graph:
\begin{eqnarray}
    \forall i \; \forall j \; \forall u \in N_i  \;
    \forall v \in N_j \;
    [\; i < j \Rightarrow u < v \;],
    \label{cond}
\end{eqnarray}
    where the ordering of the nodes is evaluated in terms of 
the labeling numbers of the nodes.
    Classes satisfying this condition are expressed as:
\begin{eqnarray}
N_1 &=& \{1, ... , |N_1|\} \nonumber\\
N_2 &=& \{|N_1| + 1, ... , |N_1| + |N_2| \} \label{kset}\\
    & & ... \nonumber
\end{eqnarray}
We consider this condition not only for the final 
classification but also classifications appearing 
in a refinement sequence.

\begin{property} \label{pr:necess}
Classification \(\{N_i^{(k)}\}\) is necessary to satisfy condition
(\ref{cond}) for \(\{N_i^{(k+1)}\}\) to satisfy the same condition.
\end{property}
\Proof{
Let \(\{N_i^{(k+1)}\}\) satisfy the condition (\ref{cond}).
A class \(N_{j}^{(k+1)}\) is expressed as eq. (\ref{kset}):
\begin{eqnarray}
N_j^{(k+1)} = \{s_j^{(k+1)}+1, s_j^{(k+1)}+2, ..., s_{j+1}^{(k+1)}\},
\label{pn:1}
\end{eqnarray}
where \(s_j^{(k+1)} = \sum_{l<j} |N_l^{(k+1)}|\).

Let a class \(N_i^{(k)}\) is decomposed to \(b_i\) classes of 
\(\{N_i^{(k+1)}\}\). 
From eq. (\ref{kclass}), we obtain,
\begin{eqnarray}
N_i^{(k)} = N_{c_i+1}^{(k+1)} \cup
            N_{c_i+2}^{(k+1)} \cup ... \cup N_{c_{i+1}}^{(k+1)},
\label{pn:2}
\end{eqnarray}
where \(c_i = \sum_{j < i} b_j\).
The following relation holds:
\begin{eqnarray}
s_{c_i+1}^{(k+1)} &=& \sum_{l < c_i +1} |N_l^{(k+1)}| \nonumber\\
    &=& s_{c_{i-1}+1}^{(k+1)} + \sum_{l=c_{i-1}+1}^{c_i} |N_l^{(k+1)}| \\
    &=& s_{c_{i-1}+1}^{(k+1)} + |N_{i-1}^{(k)}|  \nonumber \\
    &=& s_{c_{i-1}+1}^{(k+1)} + s_{i}^{(k)} - s_{i-1}^{(k)}. \nonumber
\label{pn:3}
\end{eqnarray}
Combining with \(c_1=0\) and \(s_{c_1+1}^{(k+1)} = s_1^{(k)} = 0\), 
we obtain
\begin{eqnarray}
s_{c_i+1}^{(k+1)} &=& s_i^{(k)}.
\label{pn:4}
\end{eqnarray}
Equations (\ref{pn:2}), (\ref{pn:1}) and (\ref{pn:4}) lead
\begin{eqnarray}
N_i^{(k)} &=& \cup_{j=1}^{b_i} \{s_{c_i+j}^{(k+1)} + 1,
              s_{c_i+j}^{(k+1)} + 2, ..., 
              s_{c_i+j+1}^{(k+1)}\} \nonumber\\
          &=& \{s_{c_i+1}^{(k+1)}+1, ...,  
                s_{c_{i+1}+1}^{(k+1)}\} \\
          &=& \{s_i^{(k)}+1, ..., s_{i+1}^{(k)}\}. \nonumber
\end{eqnarray}
This expression is in the form of eq. (\ref{kset}).
Thus \(\{N_i^{(k)}\}\) satisfies condition (\ref{cond})
}

    With this property, we can eliminate a graph 
before reaching to the end of the refinement sequence,
when an intermediate classification in a classification sequence 
is found not to satisfy condition (\ref{cond}).
    So we can decrease the number of refinement steps for 
graphs which are to be discarded.

    We show that graph selection using this condition (\ref{cond})
leaves enough graphs.

\begin{property} \label{pr:suf}
For each graph \(G\), there exists 
such a graph that is automorphic to \(G\) 
and satisfies condition (\ref{cond}).
\end{property}
\Proof{
    We consider the case when \(\{N_i\}\) does not satisfy
condition (\ref{cond}).
    We define a permutation \(p\) of nodes by the following
procedure:
\renewcommand{\arraystretch}{1}
\begin{quote}{\tt \begin{tabbing}
xxxx\=xxxx\=xxxx\=xxxx\=xxxx\=\kill
    void renumber()\\
    \{\\
    \>    count = 1;\\
    \>    for (i = 1; i <= ({\rm The number of classes}); i++)\{\\
    \>    \>    for all (\(v \in N_i\)) \{\\
    \>    \>    \>    ({\rm let \(p v = \)} count); \\
    \>    \>    \>    count = count + 1; \\
    \>    \>    \}\\
    \>    \}\\
    \}
\end{tabbing}}
\end{quote}
\renewcommand{\arraystretch}{2}
    It is easy to see that classification \(\{p N_i \} \) is expressed
as eq. (\ref{kset}).
    So the graph \(p G\) satisfies condition (\ref{cond})
} 

    We select graphs by the following condition instead of 
eq. (\ref{maxcode}):
\begin{eqnarray}
    f(G) = \max_{p \in S(\{N_i\})} f(p G)
                         \label{submax}
\end{eqnarray}
    Our modified orderly algorithm becomes:
\begin{itemize}
\item[1] Generate all the possible graphs.
\item[2] Select graphs by condition (\ref{cond}).
\item[3] Apply all permutations in \(S(\{N_i\})\) to each graph.
\item[4] Discard a graph when it has a smaller coding value 
         than ones of permuted graphs.
\end{itemize}

    In order to show that two selection rules (\ref{cond}) 
and (\ref{submax}) are consistent each other,
we prove the following relation for a graph \(G\) satisfying (\ref{cond}):
\begin{eqnarray}
\{pG | p \in S_{|N|}, p G \mbox{ satisfies condition (\ref{cond})}\}
    = \{ p G | p \in S(\{N_i\}) \}.
\end{eqnarray}

\begin{property} \label{pr:nec}
When nodes of a graphs \(G\) are classified into \(\{N_i\}\) 
satisfying condition (\ref{cond}) and a graph \(H = pG\) is
isomorphic to \(G\) with permutation p,
necessary and sufficient condition that a graph \(H\)
satisfies condition (\ref{cond}) is
that the permutation \(p\) is an element of 
the subgroup \(S(\{N_i\})\).
\end{property}

\Proof{
    Let the nodes of isomorphic graph \(H = p G\) be classified 
into \(\{M_j\}\).
    From eq. (\ref{iso:seq}),
a class \(N_i\) of \(G\) is mapped by \(p\) to the 
class \(M_i = p N_i\) with the same index.

    When a permutation \(p\) is an element of \(S(\{N_i\})\),
each class \(N_i\) is kept invariant under \(p\),
that is, \(N_i = p N_i = M_i\).
    Since \(\{N_i\}\) satisfies condition (\ref{cond}),
\(\{M_i\}\) satisfies the same condition.

    Inversely, we assume that the graph \(H\) satisfies the condition
(\ref{cond}).
    Since \(|N_i| = |p N_i| = |M_i|\) holds for all \(i\), it is easy
to see from expression (\ref{kset}) that \(N_i = M_i = p N_i\) is
satisfied for all \(i\).
    This implies that the permutation \(p\) keeps \(\{N_i\}\)
invariant and that \(p\) is an element of \(S(\{N_i\})\)
}

    We have first confirmed that necessary graphs are kept
in step 2 of our modified algorithm.
    Then surviving isomorphic graphs are
transformed each other by an element of subgroup \(S(\{N_i\})\).
    They are compared through the values of coding in step 4
and duplicated graphs are properly eliminated.

    In the next section we combine step 1 and 2 
in order to decrease the number of generated graphs.

\section{Graph generation and selection}
\label{sec:gen}
\newcommand{\intg}[1]{G^{[#1]}}
\newcommand{\inte}[1]{E^{[#1]}}
\newcommand{\inta}[1]{\Gamma_{G^{[#1]}}}
\newcommand{\level}{\mbox{level}}

Here we describe how to generate graphs in combination with elimination 
of graphs.

    The first step of graph generation is to prepare nodes.
    A node is to be connected to several edges.
    We consider such an imaginary object, let us call it a {\sl leg}, 
that is a place in a node where an edge is to be tied to.
    A fixed number of legs is assigned to each node at the beginning.
    The number of legs becomes equal to the degree of the node 
when connection process terminates.

    The next step is to connect nodes by edges.
    The nodes are connected iteratively starting from one fixed
external particle, which we call the {\sl root} of the graph.
    In order to make our way of connections systematic,
we define {\sl level} of node \(v\) as the distance of the node
from the root measured by the minimum number of edges
among paths connecting \(v\) and the root.
    In this step there appear intermediate configurations of
graphs, in which only a part of legs of nodes are tied with edges 
and some nodes may remain isolated from other nodes.
    We consider the level of an isolated node is infinity.

    Nodes are connected in the following way.
    The root, with one leg, is first connected to another node.
    The connected node becomes the only one node at the first level.
    Then this node is connected to other nodes until all of its legs
are connected.
    These connections define the set of nodes in the second level.
    Then, the all legs of all nodes in the second level are connected
to others.
    In this way, nodes are connected in increasing order of the
value of level.
    Once a level of a node has a finite value, its value is not
changed.
    This process proceeds in a recursive way so as to exhaust
all possibilities of the connections.
    A skeleton description of the algorithms is given as
follows:
\renewcommand{\arraystretch}{1}
\begin{quote}
{\tt \begin{tabbing}
xxxx\=xxxx\=xxxx\=xxxx\=xxxx\=\kill
    void gsconn()\\
    \{\\
    \>    node  ns, nt;\\
    \\
    \>    ns = ({\rm find node at the lowest level with free
                     legs});\\
    \>    if({\rm no more such nodes}) \{\\
    \>    \>    if({\rm the graph is a connected graph}) \{\\
    \>    \>    \>    if({\rm the graph is accepted by the orderly
                              algorithm})\\
    \>    \>    \>    \>    ({\rm a new graph is obtained});\\
    \>    \>    \}\\
    \>    \} else \{\\
    \>    \>    for all ({\rm node} nt {\rm with free legs}) \{\\
    \>    \>    \>    ({\rm connect} ns {\rm to} nt);\\
    \>    \>    \>    gsconn();\\
    \>    \>    \>    ({\rm disconnect} ns {\rm from} nt);\\
    \>    \>    \}\\
    \>    \}\\
    \}
\end{tabbing}}
\end{quote}
\renewcommand{\arraystretch}{2}

    For efficient graph generation, it is important to
eliminate unnecessary intermediate graphs in as earlier stage
as possible, since a large number of graphs may be produced
in exhausting all possible additional connection to the intermediate 
graph.
    We eliminate graphs with condition (\ref{cond})
not only in the final form of the generated graphs, but also in
intermediate graphs.

\begin{figure*}[htb]
\centerline{\epsfbox{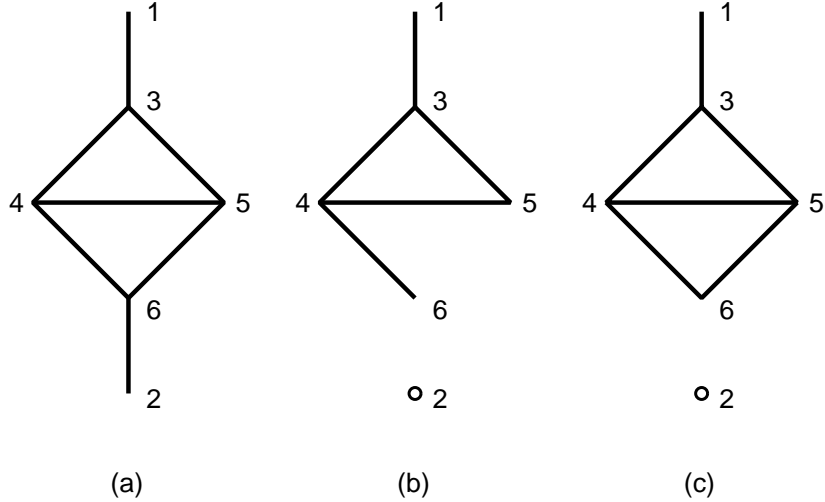}}
\caption{Example of a two-loop graph and its intermediate
         configurations.  \label{fig:twoloop}}
\end{figure*}
%

    One must notice that the classification of an
intermediate graphs is not always consistent with that of the final
form of the graphs.
    For example, we consider a final form of the graph of Fig.1a and
its intermediate graph Fig.1b.
    The set of classes of nodes of Fig.1a is
\[
    \{(1), (2), (3), (4, 5), (6)\},
\]
while that of Fig.1b is
\[
    \{(1), (2), (3), (4), (5), (6)\}.
\]
    This example shows that nodes \(4\) and \(5\) are 
topologically equivalent
in the final form of the graph, although they are not always
equivalent along the way of graph generation.
    When one applies condition (\ref{cond}) to such graphs, 
one looses necessary graphs.
    It is necessary to apply the condition (\ref{cond}) only to the 
consistent classification.

    We can recover a consistent classification by considering
an intermediate graph
\(\intg{l} = (N, \inte{l})\) at the time when all of the 
legs of all of the nodes in level \(l\) are just connected.
    These intermediate graphs form a finite series,
\[
    \intg{0} \subset \intg{1} \subset ... \subset \intg{m-1} = G,
\]
    where \(m\) is the maximum level of the nodes in 
graph \(G\).
    Fig.1c represents an example of \(\intg{2}\).
    The set of edges \(\inte{l}\) is expressed as:
\begin{eqnarray}
\inte{l} = \{ (u, v) \in E | \min( \level(u), \level(v) ) \leq l \}.
\end{eqnarray}

    We show the following property for the automorphism group
\(\inta{l}\) of \(\intg{l}\):
\begin{property}
The following relation holds for all \(l\):
\begin{eqnarray}
    \inta{l+1} \subset \inta{l}.
\label{intas}
\end{eqnarray}
\end{property}
\Proof{
    If this is not satisfied, there exists a permutation
\(p \in \inta{l+1}\) which is not an element of \(\inta{l}\).
    This means that there exists an edge \((u, v) \in \inte{l}\) and
\((p u, p v) \not\in \inte{l}\).
    On the other hand, since \(p\) is an automorphism in
\(\inta{l+1}\) and \(\level(p u) = \level(u)\),
we get:
\begin{eqnarray}
\min(\level(p u), \level(p v)) = \min(\level(u), \level(v)) \leq l,
\end{eqnarray}
which implies \((p u, p v) \in \inte{l}\)
}

    Relation (\ref{intas}) implies
\(\Gamma_G = \inta{m-1} \subset \inta{l}\).
    Thus an orbit of \(\inta{l}\) is a direct sum of the orbits of
\(\Gamma_G\).
    Since a class of nodes constructed for \(G^{[l]}\) is
a direct sum of orbits of the automorphism group \(\inta{l}\),
the class is a direct sum of the orbits of \(\Gamma_G\).
    The classification of nodes in \(\intg{l}\) is therefore 
consistent with the classification in \(G\).

    We can now eliminate irrelevant intermediate graphs with
condition (\ref{cond}) applying to \(G^{[l]}\).
    Furthermore, it is possible to use 
the classification of nodes in \(\intg{l-1}\) as an
input for constructing the classification in \(\intg{l}\)
in our class refinement algorithm. \cmmnt{???}
    This method decrease the total number of class refinement.
    The final form of our algorithms is as follows:
\renewcommand{\arraystretch}{1}
\begin{quote}
{\tt \begin{tabbing}
xxxx\=xxxx\=xxxx\=xxxx\=xxxx\=\kill
    void gsconn(int lvl)\\
    \{\\
    \>    node  ns, nt;\\
    \\
    \>    if ({\rm all legs of the nodes in the level} lvl {\rm have been}\\
    \>    ~~~~{\rm connected}) \{\\
    \>    \>    ({\rm refine classes of the nodes}); \\
    \>    \>    if(!({\rm class ordering condition (\ref{cond})
                          is satisfied})) \\
    \>    \>    \>    return;\\
    \>    \>    lvl++;\\
    \>    \}\\
    \>    ns = ({\rm find node at the level} lvl 
                {\rm with free legs});\\
    \>    if({\rm no more connectable legs}) \{\\
    \>    \>    if({\rm the graph is connected one}) \{\\
    \>    \>    \>    if({\rm the graph is accepted by the orderly
                              algorithm}) \\
    \>    \>    \>    \>    ({\rm a new graph is obtained});\\
    \>    \>    \}\\
    \>    \} else \{\\
    \>    \>    for all ({\rm node} nt {\rm with free legs}) \{\\
    \>    \>    \>    ({\rm connect} ns {\rm to} nt);\\
    \>    \>    \>    gsconn();\\
    \>    \>    \>    ({\rm disconnect} ns {\rm from} nt);\\
    \>    \>    \}\\
    \>    \}\\
    \}
\end{tabbing}}
\end{quote}
\renewcommand{\arraystretch}{2}

\section{Implementation}
\label{impl}

\prg{Particle assignment}
    We have implemented the method described above as a computer
program written in C language.
    The program generates Feynman graphs
in which particles are assigned to propagators.
    The particles of the propagators are determined in accordance
with a table of particles and interactions defined by the users.
    It is possible that duplicated graphs are produced in this
particle assignment process.
    They are eliminated again by the orderly algorithm with extended
coding of graphs with particle attributes on the edges.

\prg{Options}
    The program has the following options for graph generation:
\begin{enumerate}
\item To pick up only one-particle irreducible graphs\cite{opi}.
\item Not to generate graphs with a self-energy part at an external
      particle line.
\item Not to generate graphs with a self-energy part at an internal
      particle line.
\item To generate graphs with counter terms for renormalizing the
      theory.\\
      Counter terms are automatically generated when their
      interactions are of the same form as tree vertices.
      Other counter terms can be added by users (see appendix
      \ref{model}).
\item To generate a skeleton graphs, in which looped one-particle
      irreducible subgraphs are considered as to be blobs.
      The generated graphs are of the tree type, but with blob
      vertices.
\item Not to assign particles to the internal lines (only topology).
\end{enumerate}
    The way to specify these options is described in appendices
\ref{manual} and \ref{process}.

\begin{figure*}[htb]
\centerline{\epsfbox{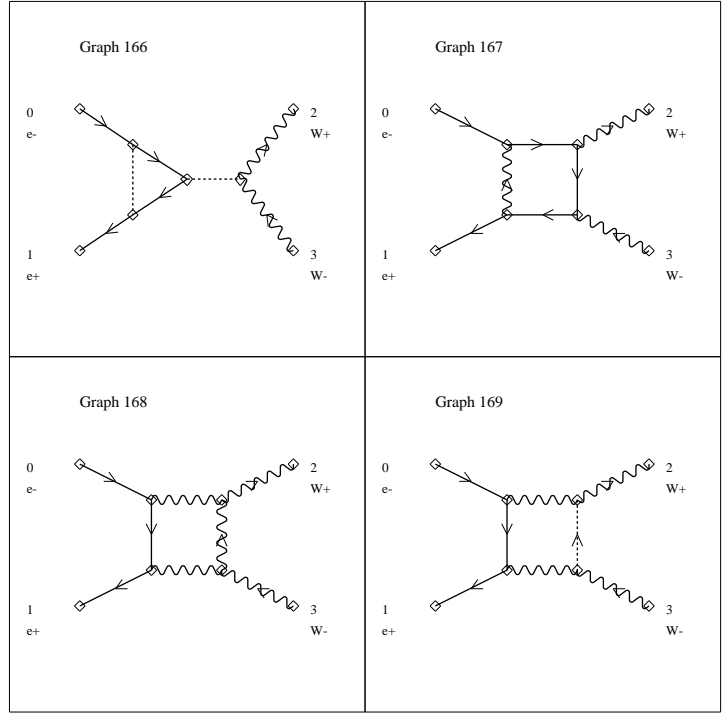}}
\caption{Example of drawn graphs.  \label{fig:plot}}
\end{figure*}

\prg{Appendices}
    We also provide a graph-drawing facility on the X-window system
and on PostScript files.
    We show an example of output figures in Fig.\ref{fig:plot}.

\prg{Running the program}
The total number of graphs depends on the conditions of the graph
selection.
    In special cases, they are analytically calculable.
    In the \(\phi^4\) model including \(\phi^3\) interaction, 
the method of enumerating connected 
graphs is developed in a graph-theoretical method \cite{read,enum}.
    When there appear different kinds of particles in the model,
a weighted sum by the symmetric factor can be calculated 
in zero-dimensional field theory \cite{zerodim}.
    These numbers of graphs are calculated and listed in
ref.\cite{nogueira}.
    We have checked our program by comparing with these numbers.
    Since for up to 5-loop graphs the orderly algorithm is not 
necessarily required to generate unique graphs \cite{fixing},
we have also checked the number of graphs for 6-loop tadpole, 
which includes 90156 graphs\footnote{The corresponding value in
\cite{fixing} is not correct.}.
    The number of one-particle irreducible graphs is
checked using a recursion formula for one- and two-loop 
graphs.

\begin{table}[htbp]
\renewcommand{\arraystretch}{1}
{\footnotesize
\begin{tabular}{||r|r|r||r|r||r|r||r|r||}
\hline\hline
A & B & C & \multicolumn{2}{c||}{D} & \multicolumn{2}{c||}{E} & \multicolumn{2}{c||}{F}\\
 &  &  & G & H & G & H & G & H\\
\hline1 & 1 & 1 & 1 & --- \qquad & 1 & --- \qquad & 1 & --- \qquad\\
1 & 2 & 6 & 15 & --- \qquad & 6 & --- \qquad & 6 & --- \qquad\\
1 & 3 & 46 & 1001 & --- \qquad & 49 & --- \qquad & 49 & --- \qquad\\
1 & 4 & 471 & 210880 & 152.1 & 570 & 152.1 & 570 & --- \qquad\\
1 & 5 & 5961 & 100368685 & 96948.8 & &  & 7451 & 16.5\\
1 & 6 & 90156 & &  & &  & 127429 & 481.9\\
\hline2 & 1 & 3 & 5 & --- \qquad & 3 & --- \qquad & 3 & --- \qquad\\
2 & 2 & 29 & 240 & --- \qquad & 29 & --- \qquad & 29 & --- \qquad\\
2 & 3 & 351 & 40302 & 25.4 & 361 & 27.0 & 361 & --- \qquad\\
2 & 4 & 5076 & 16201383 & 14280.2 & &  & 5435 & 11.2\\
2 & 5 & 84749 & &  & &  & 92739 & 291.3\\
\hline3 & 0 & 1 & 1 & --- \qquad & 1 & --- \qquad & 1 & --- \qquad\\
3 & 1 & 14 & 49 & --- \qquad & 14 & --- \qquad & 14 & --- \qquad\\
3 & 2 & 217 & 7113 & 4.0 & 217 & 4.4 & 217 & --- \qquad\\
3 & 3 & 3729 & 2486650 & 1981.2 & 3777 & 2207.8 & 3777 & 6.6\\
3 & 4 & 70600 & &  & &  & 72774 & 194.0\\
\hline4 & 0 & 4 & 7 & --- \qquad & 4 & --- \qquad & 4 & --- \qquad\\
4 & 1 & 99 & 1056 & --- \qquad & 99 & --- \qquad & 99 & --- \qquad\\
4 & 2 & 2214 & 344742 & 254.6 & 2214 & 281.6 & 2214 & 3.4\\
4 & 3 & 50051 & 217154470 & 214880.7 & &  & 50382 & 113.5\\
\hline5 & 0 & 25 & 100 & --- \qquad & 25 & --- \qquad & 25 & --- \qquad\\
5 & 1 & 947 & 38960 & 27.2 & 947 & 29.7 & 947 & 1.4\\
5 & 2 & 28365 & 24532920 & 22388.4 & &  & 28365 & 56.9\\
\hline6 & 0 & 220 & 2750 & --- \qquad & 220 & 1.3 & 220 & --- \qquad\\
6 & 1 & 11460 & 2214170 & 1862.0 & &  & 11460 & 21.3\\
\hline7 & 0 & 2485 & 121520 & 41.1 & &  & 2485 & 3.3\\
\hline\hline
\end{tabular}}
\renewcommand{\arraystretch}{2}
\caption{Efficiency of acceleration
         in the \(\phi^4\) model including \(\phi^3\) interaction.
         \label{table:gps}}
\end{table}
    The performance of the program is shown in table \ref{table:gps}
in the case of connected graphs in the same model.
Columns A, B and C represent the number of external particles,
the number of loops and the number of graphs, respectively.
    The theoretical prediction of the number of graphs is calculated
by using a program written in formula manipulate language REDUCE, 
which takes more cpu-time than our graph-generation program.
Columns D, E and F correspond to simple application of the orderly
algorithm (D), 
application of condition (\ref{cond}) without (E) and with (F)
elimination of intermediate graphs.
    Sub-columns G and H are the number of graphs applied to the orderly
algorithm eq. (\ref{maxcode}) or (\ref{submax}) 
and cpu-time in seconds measured on SUN-IPX
with ``{\tt -a -o /dev/null}'' options 
(described in appendix \ref{manual}).
This result shows that elimination of intermediate graphs
is effective enough; it reduces cpu-time in several order of magnitudes.
\section{Summary and comments}
\label{summary}

\prg{Summary}
    We have developed a method of Feynman graph generation, which
accelerates the orderly algorithm, based on a node classification
method and systematic elimination of intermediate graphs.
    This method is implemented as a computer program.
    The order of the coupling constants of a physical process is not
limited.
    Graphs can be generated based on a user-defined model; in this
paper we adopt electro-weak theory combined with QCD as an example.
    The program has several options of graph selection, including
to generate renormalization counter terms.
    Although the asymptotic behavior of the execution time is
proportional to the factorial of the number of nodes, the program
is sufficiently fast for practical use.

\prg{GRACE}
    This program is a part of the GRACE system, which automates 
computation of tree and one-loop scattering processes.
    With this system, the cross sections including one-loop corrections
in the processes
\(e^{+} e^{-} \rightarrow H Z\) and
\(e^{+} e^{-} \rightarrow t \bar{t}\),
and one-loop gluon corrections to the process
\(e^{+} e^{-} \rightarrow q \bar{q} \gamma\) are automatically
calculated\cite{gracel}.
    For the tree case, the cross sections of the scattering processes 
with up to five final particles are calculated as shown in table
\ref{table:proc}.
    The number of graphs is counted in covariant and Feynman gauges
for tree and one-loop processes, respectively.
    In both cases, the interactions among a Higgs particle and 
two light fermions are neglected.
    Integration over phase space is calculated using the integration 
package BASES\cite{bases}.

\newpage
\renewcommand{\arraystretch}{1.1}
{\footnotesize \begin{center}
\begin{tabular}{|c l|c|c|} \hline
\multicolumn{2}{|c|}{Process }& \#Graph & Reference\\ \hline
\(e^{+} e^{-}\)
    & \(\rightarrow  W^+ W^- \gamma\)            & 18
    & [\ref{gracep}a] \\ \hline
    & \(\rightarrow \nu_{e} \bar{\nu}_{e} Z^0\)  &  9
    & [\ref{gracep}b] \\ \hline
    & \(\rightarrow \nu_{e}\bar{\nu}_{e} H\)     & 2
    & [\ref{gracep}b] \\ \hline
    & \(\rightarrow e^{+} e^{-} H\)              & 2
    & [\ref{gracep}b] \\ \hline
    & \(\rightarrow Z^0 Z^0 H\)                  & 4
    & [\ref{gracep}b] \\ \hline
    & \(\rightarrow W^{+} W^{-} H\)              & 11
    & [\ref{gracep}b] \\ \hline
    & \(\rightarrow Z^0\;Z^0\;Z^0\)              & 9 
    & [\ref{gracep}b] \\ \hline 
    & \(\rightarrow W^{+} W^{-} Z^0\)            & 20 
    & [\ref{gracep}b] \\ \hline
    & \(\rightarrow t\;\bar{t}\;Z^0\)            & 9
    & [\ref{gracep}b] \\ \hline
    & \(\rightarrow t\;\bar{t}\;H\)              & 6
    & [\ref{gracep}b] \\ \hline
    & \(\rightarrow H\;H\;Z^0\)                  & 6
    & [\ref{gracep}b] \\ \hline
    & \(\rightarrow \gamma \gamma \gamma \)      & 9\({}^*\)
    & --- \\ \hline
\(\gamma \gamma\)
    & \(\rightarrow e^+ e^- Z^0\)                & 6
    & --- \\ \hline
\(\gamma e     \)
    & \(\rightarrow e W^+ W^-  \)                & 18
    & --- \\ \hline
\(g g\)
    & \(\rightarrow q\;\bar{q}\;\gamma\)         & 18
    & --- \\ \hline
\end{tabular} \par
\end{center}

\begin{center}
Table 2a Tree processes with the final 3-body
\end{center}
\begin{center}
\begin{tabular}{|c l|c|c|} \hline
\multicolumn{2}{|c|}{Process }& \#Graph & Reference\\ \hline
\(e^{+} e^{-}\)
  & \(\rightarrow \nu_e \bar{\nu}_e W^+ W^- \)
        & 60
  & [\ref{gracep}c] \\ \hline
  & \(\rightarrow \nu_{\mu (\tau)} \bar{\nu}_{\mu (\tau)} W^+ W^- \)
        & 36
  & [\ref{gracep}c] \\ \hline
  & \(\rightarrow e^+ e^- W^+ W^- \)
        & 114
  & [\ref{gracep}c] \\ \hline
  & \(\rightarrow \nu_e \bar{\nu}_e Z^0 Z^0 \)
        & 57
  & [\ref{gracep}c] \\ \hline
  & \(\rightarrow e^+ \nu_e W^- Z^0 \)
        & 88
  & [\ref{gracep}c] \\ \hline
  & \(\rightarrow e^+ e^- Z^0 Z^0 \)
        & 86
  & [\ref{gracep}c] \\ \hline
  & \(\rightarrow e^+ e^- \gamma \gamma \)
        & 80
  & [\ref{gracep}d] \\ \hline
  & \(\rightarrow \mu^+(\tau^+) \mu^-(\tau^-) \gamma \gamma \)
        & 40
  & [\ref{gracep}d] \\ \hline
  & \(\rightarrow \nu_e \bar{\nu}_e b \bar{b} \)
        & 21
  & [\ref{gracep}e] \\ \hline
  & \(\rightarrow  \nu_{\mu (\tau)} \bar{\nu}_{\mu (\tau)}
                   b \bar{b} \) &11
  & [\ref{gracep}e] \\ \hline
  & \(\rightarrow e^+ \nu_e \bar{t} b \)
          & 21
  & [\ref{gracep}f] \\ \hline
  & \(\rightarrow W^+ W^- \gamma \gamma  \)
          & 138
  & --- \\ \hline
  & \(\rightarrow e^- \nu_e u \bar{d}  \)
          & 24
  & [\ref{gracep}g] \\ \hline
  & \(\rightarrow u \bar{d} \bar{u} d  \)
          & 69
  & [\ref{gracep}g] \\ \hline
  & \(\rightarrow q \bar{q} g  \gamma \)
          & 21
  & --- \\ \hline
  & \(\rightarrow e^+ e^- b \bar{b}\)
          & 50
  & --- \\ \hline
\(\gamma e \)
  & \(\rightarrow \nu_e W^- H H \)
          & 40
  & --- \\ \hline
\end{tabular}\par
\end{center}

\begin{center}
Table 2b Tree processes with the final 4-body
\end{center}
\begin{center}
\begin{tabular}{|c l|c|c|} \hline
\multicolumn{2}{|c|}{Process }& \#Graph & Reference\\ \hline
\(e^{+} e^{-}\)
    & \(\rightarrow \mu^+ \nu_{\mu} \bar{t} b \gamma \) & 71
    & --- \\ \hline
    & \(\rightarrow e^- \bar{\nu}_e u \bar{d} \gamma \) & 142
    & [\ref{gracep}h] \\ \hline
\end{tabular}\par
\end{center}

\begin{center}
Table 2c Tree processes with the final 5-body
\end{center}
\begin{center}
\begin{tabular}{|c l|c|c|} \hline
\multicolumn{2}{|c|}{Process }& \#Graph & Reference\\ \hline
\(e^{+} e^{-}\)
    & \(\rightarrow H Z\)       & 89 & \cite{gracel} \\ \hline
    & \(\rightarrow t \bar{t}\) & 50 & \cite{gracel} \\ \hline
\end{tabular}\par
\end{center}

\begin{center}
Table 2d One-loop processes with the final 2-body
\end{center}
\begin{center}
\begin{tabular}{|c l|c|c|} \hline
\multicolumn{2}{|c|}{Process }& \#Graph & Reference\\ \hline
\(e^{+} e^{-}\)
    & \(\rightarrow q \bar{q} \gamma\) & 12 & \cite{gracel} \\ \hline
\end{tabular}\par
Gluon correction with only final radiations.
\end{center}

\begin{center}
Table 2e One-loop processes with the final 3-body
\end{center}
}
\renewcommand{\arraystretch}{2}
\begin{table}[h]
\caption{Scattering processes calculated by
         the GRACE system.\label{table:proc}}
\end{table}

\newpage

\vspace{1em}

\prg{Loosen conditions of graph generation}
    In the following
    we discuss how to loosen the restrictions for graph generation
assumed in the introduction.
    The second restriction regarding identical external particles
can be removed by changing the primitive classification so as to 
put identical particles in the same class.
    At the same time, the root node should be replaced by a class
of nodes.

\prg{Vacuum graphs}
    The first restriction of excluding vacuum-to-vacuum graphs
is complicated to loosen.
    It is possible that the primitive classification contains
only one class, such as in vacuum-to-vacuum graphs in the \(\phi^3\)
model.
    In this case, node classification and acceleration methods
do not work anymore.
    However, one can artificially fix one node as the root
of the graph.

\prg{Modification of generation method}
    We consider a set of graphs \(A\) generated by our method
with fixing node \(1\) among \(n = |N|\) nodes and
selected by condition (\ref{cond}).
    Let us also consider the true set of vacuum-to-vacuum graphs
\(T\).
    For any graph \(G \in T\), one can construct a set of
isomorphic graphs
\(I_G = \{ p G \,|\, p \in S_{n-1}\}\), where \(S_{n-1}\)
is the symmetric group acting on the set of the nodes
\(\{2, \cdots, n\}\).
    We consider to select a representative graph from \(I_G\) by applying 
condition (\ref{cond}) after classifying the nodes with fixing
node \(1\). 
    The selected graph belongs also to \(A\), since
\(A\) is constructed by the same selection rule applied to all
possible graphs.
    The set of graphs \(A\) thus contains at least an isomorphic 
graph to any graph of \(T\), though some graphs in \(A\) are 
duplicated.

\prg{Modification of acceleration method}
    The duplication of graphs is to be tested by the original orderly
algorithm (\ref{maxcode}).
    Since we have already eliminated some of the duplicated graphs
by condition (\ref{cond}), we must limit the set of
graphs \(\{p G \,|\, G \in A,\; p\in P\}\) by the same condition
before comparing the values of the coding.
    The orderly algorithm is then changed to the following form:
\begin{quote}
\renewcommand{\arraystretch}{1}
{\tt \begin{tabbing}
xxxx\=xxxx\=xxxx\=xxxx\=xxxx\=\kill
compare(\(G\)) \\
\{
\>   accept = True; \\
\>   for all (\(p G,\; p \in S_{|N|}\)) \{ \\
\>   \>   \(H\) = \(p G\); \\
\>   \>   ({\rm classify node of \(H\) with fixing node \(1\)}); \\
\>   \>   if({\rm condition (\ref{cond}) is satisfied}) \{ \\
\>   \>   \>  if(\(coding(G)\) < \(coding(H)\)) \{\\
\>   \>   \>  \>  accept = False; \\
\>   \>   \>  \>  break; \\
\>   \>   \} \\
\>   \} \\
\>   if(accept) \\
\>   \>   ({\rm accept graph \(G\)});\\
\}
\end{tabbing}}
\renewcommand{\arraystretch}{2}
\end{quote}

\section{Acknowledgment}
    The author wishes to thank the members of Minami-Tateya
collaboration for continuous discussions and many kinds of support.
    Especially, he owes special thanks to Dr. Y.Shimizu, whose
influence appears as the basis of this work, for his useful
suggestions and to Dr. S.Kawabata
who provided the first version of the driver program of X-Library.

    The author is also indebted for useful discussions to
Dr. P.Nogueira concerning the orderly algorithm and to 
Dr. J.A.M.Vermaseren concerning the format of the model definition 
and output files.

    This work was supported in part by Ministry of Education,
Science and Culture, Japan under Grant-in-Aid for International
Scientific Research Program (No.04044158).

\newpage
\appendix
\section{Installation and execution}
\label{manual}

    The source code is available by {\tt anonymous ftp} from
{\tt ftp.kek.jp} in the directory {\tt kek/minami/grc}.
    It has two programs: one generates Feynman graphs; the other
draws generated graphs, which work on UNIX with the X-Window system.

    The graph-generation program {\tt grc} reads two files,
a process file and a model file.
    It then creates an output file named ``{\tt out.grf}'', 
which keeps the information about the generated graphs.
    The format of the input files is described in appendices
\ref{process} and \ref{model}.

    The graph-drawing program {\tt grcdraw} reads the model and 
the output file of the generation program, and then draws 
the graphs on the display with the X-Window system.
    This program is not so intelligent as to display graphic objects
beautifully; however, it is still useful for checking the generated
graphs.
    This program has several subcommands.

    The procedure of installation is:
\begin{enumerate}
\item Uncompress and expand the source code file.\\
      In a new directory, run the following command:
      \begin{center}
      {\tt zcat grc.tar.Z | tar xvf - }
      \end{center}

\item Editing {\tt Makefile}.\\
      The file {\tt Makefile} in the sub-directory {\tt src} controls
      the way of compiling source code.
      The path-names of include files and libraries of {\tt Xlib}
      should be changed in accordance with the system configuration.
      If {\sl Athena widgets} (included in the standard distribution
      of X-Window system) is installed in your system, use the line
      in the {\tt Makefile}:
      \begin{center}
      {\tt XCOPT     = -DTOOLKIT -DX11}
      \end{center}
      If not, use
      \begin{center}
      {\tt XCOPT     = -DX11}
      \end{center}

\item Compilation.\\
      In the sub-directory {\tt src}, run {\tt make} command.

\item Testing the Feynman-graph generator.\\
      In the sub-directory {\tt sample}, run the following command
      \begin{center}
      {\tt ../src/grc}
      \end{center}
      The program generates 4 tree graphs for
      the \(e^+ e^- \rightarrow W^+ W^-\) process, 316 one-loop graphs
      of the same process and 28 tree graphs for
      \(e^+ e^- \rightarrow W^+ W^- \gamma\).

\item Testing the Feynman-graph drawer.\\
      In the same sub-directory {\tt sample}, run the following
      command
      \begin{verse}
      {\tt ../src/grcdraw -h} \\
      {\tt ../src/grcdraw}
      \end{verse}
      Try to type ``{\tt f}'', ``{\tt b}'', ``{\tt n}'', ``{\tt p}'',
      ``{\tt g}'' ``{\tt q}''.

      If you have installed with Athena widgets, buttons will
      appear on the display.
\end{enumerate}

\vspace{1em}
\leftline{\large\bf Feynman-graph generator}
\vspace{0.5em}

The command syntax of {\tt grc} command is
\begin{center}
{\tt grc [{\sl options}] [{\sl process-file-name}]}
\end{center}
The {\sl process-file-name} is an input file specifying
the physical process and options, the format of which is described
in appendix \ref{process}.
The default process file name is ``{\tt in.prc}''.

The command line options of the {\tt grc} command are:
\begin{itemize}
\item {\tt -a} \\
      Skip particle assignment.
\item {\tt -c} \\
      Do not generate an output file, but only count the number of graphs.
\item {\tt -o} {\sl output-file-name} \\
      Specify the output file name.
      The default output file is ``{\tt out.grf}''.
\item {\tt -h}\\
      Print help message.
\end{itemize}
Also try with the help option:
\begin{center}
{\tt ../src/grc -h}
\end{center}

\vspace{1em}
\leftline{\large\bf Feynman-graph drawer}
\vspace{0.5em}

The command syntax of the {\tt grcdraw} command is
\begin{center}
{\tt grcdraw [{\sl options}] [{\sl graph-file-name}]}
\end{center}
The {\sl graph-file-name} is an output file generated by the {\tt grc}
command, the format of which is described in appendix \ref{grf}.
The default process file name is ``{\tt out.grf}''.

The command line options of the {\tt grcdraw} command are:
\begin{itemize}
\item {\tt -w} {\sl number-of-graphs}\\
      If this option is specified
      ({\sl number-of-graphs})\({}^2\) graphs appear
      on the display.
\item {\tt -h}\\
      Print help message.
\end{itemize}
Also try with the help option:
\begin{center}
{\tt ../src/grcdraw -h}
\end{center}

This program has some subcommands for selecting graphs drawn on the
display.
They are as follows:
\renewcommand{\arraystretch}{1}
\begin{center} \begin{tabular}{rl}
    {\tt q} :& quit.\\
    {\tt n} :& display next process \\
    {\tt p} :& display previous process \\
    {\tt f} :& display forward graphs \\
    {\tt b} :& display backward graphs \\
    {\tt g} :& scale up the size of graphs \\
    {\tt s} :& scale down the size of graphs \\
    {\tt l} :& display particle names of internal lines (on/off)\\
    {\tt t} :& display particle names and graph number (on/off)\\
    {\sl \lbni number\rbni}{\tt j} :& jump to the specified graph\\
    {\tt o} :& output the displayed graphs to PostScript file
\end{tabular} \end{center}
\renewcommand{\arraystretch}{2}

When the program is installed with Athena widgets,
such commands are also displayed as buttons.
\section{Process file}
\label{process}

    The graph-generation program works in accordance with the process
file.
    The following example is a file specifying the one-loop and tree
process of \(e^+ e^- \rightarrow W^+ W^-\) and the tree process of
\(e^+ e^- \rightarrow W^+ W^- \gamma\).

\begin{quote}
{\footnotesize\begin{verbatim}
%%%%%%%%%%%%%%%%%%%%%%%%%%%%%%%%%%%%%%%%
Model="all.mdl";
%%%%%%%%%%%%%%%%%%%%%%%%%%%%%%%%%%%%%%%%
Process;
  ELWK={2, 4};
  Initial={electron, positron};
  Final  ={W-plus, W-minus};
  Expand=Yes;
  OPI=No;
Pend;
Process;
  ELWK=3;
  Initial={electron, positron};
  Final  ={W-plus, W-minus,photon};
  Expand=Yes;
  OPI=No;
Pend;
\end{verbatim}}
\end{quote}

    The lines beginning with ``\verb+%+'' are ignored as comment
lines.

    The first non-comment line specifies the file which describes
the model used for graph generation, which is described in appendix
\ref{model}.

    Then follow descriptions of the processes.
    A block describing a process begins with the the line of 
``{\tt Process;}'' and ends with the line of ``{\tt Pend;}''.
    In this block, descriptions are given for the order of 
the coupling constants, initial particles, final particles and some 
options.

    The name of the coupling constant (``{\tt ELWK}'' in the above
example) is defined in the model file.
    When the value of the coupling constant is given as a list of
numbers, the program generates graphs for each value of the coupling
constants with the same external particles.

    The initial and final external particles are given as lists of
particle names, which are defined in the model file.

    The available options are :
\begin{itemize}
\item {\tt OPI = Yes | No}\\
      Generate one-particle irreducible graphs or not.
\item {\tt Expand = Yes | No}\\
      Expand the looped part or generate looped parts as blobs.
\item {\tt Tadpole = Yes | No}\\
      Generate tadpoles or not.
\item {\tt extself = Yes | No}\\
      Generate the self-energy part at an external particle or not.
\item {\tt selfe = Yes | No}\\
      Generate whether with the self-energy part or not at an internal
      particle line.
\item {\tt countert = Yes | No}\\
      Generate renormalization counter terms or not.
\end{itemize}
\newpage
\section{Model file}
\label{model}

    Here, we describe the format of the model file.
    Our definitions of particles and vertices include not only
the necessary information for graph generation, but also for amplitude
generation.
    We show an example of this file:
\begin{quote}
{\footnotesize\begin{verbatim}
%%%%%%%%%%%%%%%%%%%%%%%%%%%%%%%%%%%%%%%%
 Order={ELWK, QCD};
%=======================================
% gauge bosons
%---------------------------------------
 Particle=W-plus["W+"]; Antiparticle=W-minus["W-"];
     PType=Vector; Charge=1; Color=1; Mass=AMW; Width=AGW;
     MValue="80.22D0"; WValue="2.12D0"; PCode=2;
 Pend;

    ...

%---------------------------------------
% scalars
%---------------------------------------
 Particle=Higgs["H"]; Antiparticle=Particle;
     PType=Scalar; Charge=0; Color=1; Mass=AMH; Width=AGH;
     MValue="150.0D0"; WValue="0.0D0"; PCode=31;
 Pend;
%
         ...

%---------------------------------------
% leptons
%---------------------------------------
 Particle=nu-e["nue"]; Antiparticle=nu-e-bar["~nue"];
     PType=Fermion; Charge=0; Color=1; Mass=AMNE; Width=0;
     MValue="0.0D0"; PCode=51; Massless;
 Pend;

    ...

 Particle=electron["e-"]; Antiparticle=positron["e+"];
     PType=Fermion; Charge=-1; Color=1; Mass=AMEL; Width=0;
     MValue="0.511D-3"; PCode=55;
 Pend;

    ...

%---------------------------------------
% quarks
%---------------------------------------
 Particle=u; Antiparticle=u-bar["~u"];
     PType=Fermion; Charge=2/3; Color=3; Mass=AMUQ; Width=0;
     MValue="100.0D-3"; PCode=61;
 Pend;

   ...

%=======================================
% VVV
%---------------------------------------
 Vertex={Z,      W-minus, W-plus}; ELWK=1; FName=ZWW;
        FValue={R, " GG*GCOS"}; Vend;
 Vertex={photon, W-minus, W-plus}; ELWK=1; FName=AWW;
        FValue={R, " GE"}; Vend;
 Vertex={gluon,  gluon,  gluon  };  QCD=1; FName=GGG;
        FValue={R, " CQCD"}; Vend;
%---------------------------------------
% VVVV
%---------------------------------------
 Vertex={W-plus, W-minus, photon,  photon }; ELWK=2;
        FName=WWAA; FValue={R, " GE2"}; Vend;
 Vertex={W-plus, W-minus, Z,       photon }; ELWK=2;
        FName=WWZA; FValue={R, " GE*GG*GCOS"}; Vend;

    ...

%---------------------------------------
% FFV (FFW : without quark mixing)
%---------------------------------------
 Vertex={positron,  nu-e,      W-minus}; ELWK=1; FName=WNE;
        FValue={R, " GWFL"}; FType="V-A"; Vend;
 Vertex={anti-muon, nu-mu,     W-minus}; ELWK=1; FName=WNM;
        FValue={R, " GWFL"}; FType="V-A"; Vend;
 Vertex={anti-tau,  nu-tau,    W-minus}; ELWK=1; FName=WNT;
        FValue={R, " GWFL"}; FType="V-A"; Vend;
 Vertex={nu-e-bar,  electron,  W-plus }; ELWK=1; FName=WEL;
        FValue={R, " GWFL"}; FType="V-A"; Vend;
 Vertex={nu-mu-bar, muon,      W-plus }; ELWK=1; FName=WMU;
        FValue={R, " GWFL"}; FType="V-A"; Vend;
 Vertex={nu-tau-bar,tau,       W-plus }; ELWK=1; FName=WTA;

    ...

%---------------------------------------
% Counter terms introduced by photon-Z mixing.
%---------------------------------------
% SSV
%---------------------------------------
 Vertex={chi-3,     Higgs,    photon }; ELWK=3; FName=AHY;
        Vend;
%---------------------------------------
% SVV
%---------------------------------------
 Vertex={Higgs,     Z,      photon  }; ELWK=3; FName=HZA;
        Vend;
 Vertex={Higgs,     photon, photon  }; ELWK=3; FName=HAA;
        Vend;

    ...

%---------------------------------------
% FFV (FFZ)
%---------------------------------------
 Vertex={nu-e-bar,   nu-e,     photon}; ELWK=3; FName=ANE;
        Vend;
 Vertex={nu-mu-bar,  nu-mu,    photon}; ELWK=3; FName=ANM;
        Vend;
 Vertex={nu-tau-bar, nu-tau,   photon}; ELWK=3; FName=ANT;
        Vend;
%=======================================
Mend;
%=======================================
\end{verbatim}}
\end{quote}

    Lines beginning with ``\verb+%+'' are comment lines.

    The file comprises of three parts: definitions of the names of
the coupling constants, definitions of the particles and then
definitions of the vertices and counter-terms, arranged in this order.
    A definition is a sequence of
``{\sl keyword}{\tt =}{\sl value}{\tt ;}''
or ``{\sl keyword{\tt ;}}''.
    The {\sl value} part may have an optional part, which is enclosed
by brackets ''{\tt [}'' and ``{\tt ]}''.

    The first keyword is {\tt Order}, which define the names of the
coupling constants.
    When multiple coupling constants appear in the model,
a list of these names is specified enclosed by braces ``\verb+{+''
and ``\verb+}+''.
    For example,
\begin{center}{\tt
Order=\{QED, QCD\};
}\end{center}
is for a model including QED and QCD coupling constants.

\subsection{Definition of particles}

    The definition of a particle begins with the keyword {\tt Particle}
and ends with the keyword {\tt Pend}.

    The value of the keyword {\tt Particle} is the name of the
particle.
    The name, beginning with an alphabet, should be unique among
the particles defined in this file, since this name is used to identify
the particle.
    An option to the name is a shorthand name of the particle, which is
used to show the particle in the graphic output.

    In the same way, the name of an anti-particle is given as the value
of the keyword {\tt Antiparticle}.
    When the anti-particle coincides to the particle, this part
should be defined as
\begin{center}
{\tt Antiparticle=Particle}.
\end{center}

    The type of particle is given by the keyword {\tt PType}.
    Its value is either {\tt Scalar}, {\tt Vector}, {\tt Majorana},
{\tt Fermion} or {\tt Ghost}.

    The keyword {\tt Charge} specifies the electric charge.
The value is given as a signed integer or a rational number
in the unit of the positron charge \(e\).

    The dimension of color representation of the particle is
specified by the {\tt Color} keyword as an integer.

    The definition of the mass parameter of the particle comprises
three items: the Fortran variable name of the mass,
the default numerical value and a flag specifying massive or massless.
    Even if the particle is massless, a fictitious mass can be
introduced in some part of the calculation in order to avoid an infrared
divergence or mass singularity.
    We require a definition of a particle being massless or massive,
and the Fortran variable name for the particle mass with a default value
not only for a massive particle, but also a massless one.

    The keyword {\tt Massless} or {\tt Massive} (without a value)
specifies that the particle is either massless or massive, respectively.
    If nothing is specified, the particle is considered to be massive.

    The Fortran name is defined by the keyword {\tt Mass} and its
default value by {\tt MValue}.
    The value of keyword {\tt MValue} is defined as a character
string which is used in the Fortran code.

    The Fortran name of the width and its default numerical value is
given by the keywords {\tt Width} and {\tt WValue}, respectively.

    In the above definition, we do not assume any special name of
particles; one can define the particle name freely.
    However, amplitude-generation program is necessary to know
whether a special particle appears in a Feynman graph or not.
    For example, the {\tt CHANEL} library offers a calculation of
the amplitudes in a general covariant gauge.
    In the unitary gauge, \(\chi\)-scalars disappear form
the calculation, and one must drop any Feynman graphs including them.
    In order to detect such kinds of particles, particularly
in the amplitude generating-program, we add another keyword
{\tt PCode} with an integer particle code.
    The values of {\tt PCode} are not used in the Feynman-graph
generator in the current version.

\subsection{Definition of vertices}

    A definition of a particle begins with the keyword {\tt Vertex}
and ends with the keyword {\tt Vend}.

    The keyword {\tt Vertex} defines the interacting particles in the
list of their name.
    The direction of a particle is defined as incoming to the vertex.
    For example,
\begin{center} {\tt
 Vertex=\{nu-e-bar,  electron,  W-plus\};
}\end{center}
defines a vertex which \(e^-\) and \(W^+\) are coming in and \(\nu_e\)
is going out.

    The order of the coupling constants is given by keywords defined
in the value of the keyword {\tt Order} such as:
\begin{center} {\tt
ELWK=1;
}\end{center}
    The total order of coupling constant of tree 3- and 4-point 
vertices should be 1 and 2, respectively.

    A program generating amplitudes requires additional
information.
    The name of the coupling constant used in the Fortran code 
of the vertex is given by the {\tt FName} keyword, whose data type 
and default value are given by the {\tt FValue} keyword.
    The number of parameters in a coupling constants of a vertex depends on 
the interaction.
    In order to decide its number and how particles interact in
a vertex of two fermions and a boson, the type of the interaction
is specified by the keyword {\tt FType}.
    The value of {\tt FType} is either {\tt "V"}, {\tt "A"}, 
{\tt "S"}, {\tt "P"}, {\tt "V-A"}, {\tt "V+A"}, {\tt "S-P"}, 
{\tt "S+P"} or {\tt "NON"}.
    When this keyword is omitted, the fermion-boson vertex is
assumed to {\tt FType="NON"}.
    The format of specification by the {\tt FValue} keyword is
\begin{center}
{\tt FValue=\{}{\sl \lbni flag\rbni, \lbni param1\rbni}{\tt\};}
\end{center}
or
\begin{center}
{\tt FValue=\{}{\sl \lbni flag\rbni, \lbni param1\rbni, 
                    \lbni param2\rbni}{\tt\};}
\end{center}
in accordance with the number of coupling constants.
The parts {\sl \lbni param1\rbni} and {\sl \lbni param2\rbni} 
are embedded into the Fortran code by the Fortran-code generator.
The part {\sl \lbni flag\rbni} is either {\tt R}, {\tt I} or {\tt C},
corresponding to the parameter being real, pure imaginary or complex
number, respectively.
    These keywords are not used in the graph-generation program.

    One can define a counter term of renormalization as a vertex
with higher order coupling constants.

\newpage
\section{Output file}
\label{grf}

    We show an example of the output file of the graph generator
for the input file shown in appendix \ref{process}.
\begin{quote}
{\footnotesize\begin{verbatim}
%%%%%%%%%%%%%%%%%%%%%%%%%%%%%%%%%%%%%%%%
Model="all.mdl";
%%%%%%%%%%%%%%%%%%%%%%%%%%%%%%%%%%%%%%%%
Process=1;
External=4;
   0= initial electron;
   1= initial positron;
   2= final w-plus;
   3= final w-minus;
Eend;
elwk=4;Loop=1;
OPI=No;Expand=Yes;
%---------------------------------------
Graph=1;
Sfactor=1;
Vertex=4;
   0={    1[positron]};
   1={    2[electron]};
   2={    3[w-plus]};
   3={    4[w-minus]};
   4[order={1,0}]={  1[electron],  2[positron],  5[photon]};
   5[order={1,0}]={  5[photon],    6[w-minus],   7[w-plus]};
   6[order={1,0}]={  3[w-minus],   6[w-plus],    8[z]};
   7[order={1,0}]={  4[w-plus],    7[w-minus],   8[z]};
Vend;
Gend;
%---------------------------------------
Graph=2;

    ...

%---------------------------------------
Graph=316;
Sfactor=1;
Vertex=3;
   0={    1[positron]};
   1={    2[electron]};
   2={    3[w-plus]};
   3={    4[w-minus]};
   4[order={1,0}]={  1[electron],  4[w-plus],   5[nu-e-bar]};
   5[order={1,0}]={  2[positron],  3[w-minus],  6[nu-e]};
   6[loop=1;order={2,0}]={  5[nu-e],  6[nu-e-bar]};
Vend;
Gend;
%---------------------------------------
Pend=316;
%%%%%%%%%%%%%%%%%%%%%%%%%%%%%%%%%%%%%%%%
Process=2;
External=4;
   0= initial electron;
   1= initial positron;
   2= final w-plus;
   3= final w-minus;
Eend;
elwk=2;Loop=0;
OPI=No;Expand=Yes;
%---------------------------------------
Graph=1;

    ...

Graph=28;
Sfactor=1;
Vertex=2;
   0={    1[positron]};
   1={    2[electron]};
   2={    3[w-plus]};
   3={    4[w-minus]};
   4={    5[photon]};
   5[order={1,0}]={  1[electron],  2[positron],  6[z]};
   6[order={2,0}]={  3[w-minus],   4[w-plus],    5[photon],  6[z]};
Vend;
Gend;
%---------------------------------------
Pend=28;
%%%%%%%%%%%%%%%%%%%%%%%%%%%%%%%%%%%%%%%%
End=3;
\end{verbatim}}
\end{quote}

    Lines beginning with ``\verb+%+'' are ignored as comments.

    The output file is a sequence of statements.
    Most statements are written in the format
\begin{quote}
    ``{\sl keyword}{\tt =}{\sl value}'' \quad or \quad
``{\sl keyword}''.
\end{quote}
    separated by colons ``{\tt ;}''.
    The {\sl keyword} part is an identifier (name composed of
alphabet and digits).
    The {\sl value} part is either an identifier or a list of
identifiers separated by ``{\tt ,}'' and enclosed by braces
``{\tt \{}'' and ``{\tt \}}''.
    An identifier may have options enclosed by brackets
``{\tt [}'' and ``{\tt ]}''.
    Between two brackets, a statement or a list of statements is
placed.

    The fist non-comment line specifies the file name of the model
definition which is used in graph generation.

    The output file includes several processes in accordance
with the input file.
    In this example, the output file includes three processes.
    Each of them begins with the line
\begin{quote}
     {\tt Process=}{\sl \lbni process number\rbni}{\tt ;}
\end{quote}
    and ends with the line
\begin{quote}
     {\tt Pend=}{\sl \lbni the number of graph
                     in the process\rbni}{\tt ;}
\end{quote}

    The description of a process comprises descriptions of
external particles, options for graph generation copied from the
input file and a sequence of descriptions of the generated graphs.

The description of external particles begins with the statement
\begin{quote}
   {\tt External=}{\sl \lbni the number of external
                       particles\rbni}{\tt ;}
\end{quote}
and ends with the statement
\begin{quote}
   {\tt Eend;}
\end{quote}
Between them, statements for each external particle are written
in the format
\begin{quote}
   {\sl \lbni external particle number\rbni}{\tt = initial}
   {\sl \lbni particle name\rbni} {\tt ;}
\end{quote}
or
\begin{quote}
   {\sl \lbni external particle number\rbni}{\tt = final}
   {\sl \lbni particle name\rbni} {\tt ;}
\end{quote}
The {\sl \lbni external particle number\rbni} is an integer number
beginning from 0.

The description of a generated graph begins with the statement
\begin{quote}
   {\tt Graph=}{\sl \lbni graph number\rbni}{\tt ;}
\end{quote}
and ends with the statement
\begin{quote}
   {\tt Gend;}
\end{quote}
After the {\tt Graph} statement, it follows the global factor
of the graph,
\begin{quote}
   {\tt Sfactor=}{\sl \lbni inverse of global factor\rbni}{\tt ;}
\end{quote}
The value of {\tt Sfactor} is the inverse of the symmetric
factor of the graph with a sign corresponding to the permutation of
the external fermions and fermion loops.

    The structure of a graph is expressed by connections between
the nodes.
    They are placed between two statements:
\begin{quote}
   {\tt Vertex=}{\sl\lbni the number of generated
                    vertices\rbni}{\tt ;}
\end{quote}
    and
\begin{quote}
   {\tt Vend;}
\end{quote}
    Corresponding to each node, information about the connections
from the node to others is written as
\begin{center}
\renewcommand{\arraystretch}{1}
\begin{tabular}{rl}
   {\sl \lbni node number\rbni}{\tt =\tt \{ }
        & {\sl \lbni internal line number\rbni}
    {\tt [}{\sl particle name}{\tt ], } \\
        & {\tt ... \};}
\end{tabular}
\renewcommand{\arraystretch}{2}
\end{center}
    The {\sl \lbni node number\rbni} is sequentially
numbered beginning from 0.
    A node number corresponding to one defined as
{\sl \lbni node number\rbni} expresses the same external particle.
    The {\sl \lbni internal line number\rbni} is numbered
sequentially beginning from 1.
    Two nodes with common {\sl \lbni internal line number\rbni}
are connected by the internal line.
    The {\sl \lbni particle name\rbni} is the name of a particle
assigned to the internal line defined as incoming to the node.
    When multiple coupling constants are defined in the model,
the following option is added to the {\sl \lbni node number\rbni}
\begin{quote}
    {\tt order=\{}{\sl order-1, order-2, ..., order-n\}}{\rm ,}
\end{quote}
which specifies the orders of coupling constants of the vertex,
corresponding to the {\tt Order} statement in the model file.
    Moreover, when the vertex is a blob or a counter term,
the following option is also added:
\begin{quote}
    {\tt loop=}{\sl \lbni the number of loops in the vertex\rbni}
\end{quote}

    The file ends with the {\tt End} statement:
\begin{quote}
     {\tt End=}{\sl \lbni the number of processes\rbni}{\tt ;}
\end{quote}

\end{document}